\documentclass{aa}
\usepackage{graphicx}        

\begin{document}

%
   \headnote{Research Note} 
   \title{Interaction between a galactic disk and a live dark halo \\ with an
   anisotropic velocity distribution} 

   \author{B. Fuchs\inst{1} \and E. Athanassoula\inst{2}}


   \institute{Astronomisches Rechen--Institut,
              M\"onchhofstrasse 12--14, 69120 Heidelberg, Germany
	      \and Observatoire de Marseille, 2 Place Le Verrier, 13248 
	      Marseille cedex 04, France}

   \date{Received 2005; accepted}

   \abstract{
   We have extended previous analytical studies of the interaction of
   dark halos with 
   galactic disks by introducing for the halo particles anisotropic 
   distribution functions in phase space. For this purpose we have employed
   the shearing sheet model of a patch of a galactic disk embedded in a 
   homogeneous halo. We find that velocity anisotropy increases
   considerably the maximum growth factor of perturbations in the disk.
      }
		
   \mail{fuchs@ari.uni-heidelberg.de}
   
   \titlerunning{Anisotropic live dark halos}
		
   \maketitle
%

\section{Introduction}

The dynamical evolution of non--axisymmetric structures such
as spirals or bars  is driven by the redistribution of
angular momentum within the disk galaxy that harbours them. This was
argued and demonstrated by Lynden-Bell \& 
Kalnajs (1972), who showed that stars at inner Lindblad resonance and
at higher order resonances within corotation emit angular momentum,
which is absorbed by stars at corotation, at outer Lindblad resonance and
at higher order resonances outside corotation. Non--resonant stars can
also contribute to this exchange, provided the spiral or bar
perturbation has a non--negligible growth rate. Since the non--axisymmetric 
structures are inside the corotation circle negative angular momentum 
`perturbations' (Kalnajs 1971, 
Lynden-Bell \& Kalnajs 1972), this exchange induces their growth. 
In this early picture the role of the spheroid is not included. 
However, recent, state--of--the--art $N$-body simulations have clearly
shown that it can be predominant in galaxies with strong bars. Indeed,
if the galactic disk is embedded in a spheroid or dark halo, a similar
angular momentum exchange between the bar and the
surrounding bulge or dark halo is expected. Athanassoula (2002, 2003)
studied in particular the interaction of a bar in the disk and a live 
spheroidal component (bulge and/or halo). Provided the latter has a 
distribution function that is a function of the energy only, it is possible to 
show analytically that it can at all its resonances only absorb angular
momentum. Thus in the presence of spheroid or a halo a bar can grow 
stronger than
in its absence, since in the former case more angular momentum is
taken from the bar region. Tremaine \& Weinberg (1984) and Weinberg
(1985) calculated that the angular momentum exchange also causes a
considerable slow--down of the bar, which is similarly enhanced in the
presence of a live spheroid with an isotropic distribution function.
This was, at least qualitatively, confirmed by a number of $N$--body
simulations (e.g. Little \& Carlberg 1991, Hernquist \& Weinberg 1992,
Athanassoula 1996, Debattista \& Sellwood 2000, Athanassoula 2003,
O'Neill \& Dubinski 2003, Valenzuela \& Klypin 2003).
Athanassoula (2002, 2003) used $N$--body simulations to show that a
considerable amount of halo material is at resonance with the bar. She
further traced the angular momentum exchange in the simulations
and found very good qualitative agreement with analytical predictions.
Angular momentum is emitted by particles within corotation: 
mainly by particles at inner Lindblad resonance, but also by particles
at higher order resonances and also by non--resonant particles.
It is then absorbed by disk particles at corotation and
outer Lindblad resonance, halo resonant particles at all resonances and
non--resonant halo particles. Indeed, bars in simulations with a responsive
halo, that can take part in the angular momentum exchange, grow
considerably stronger than bars in simulations with a halo which has the same 
mass distribution, but is rigid (Athanassoula 2002). This justifies, in
retrospect, the doubts voiced much earlier by Toomre (1977) as to the entirely
passive role of the halo in bar formation and development.

The dynamics of bars and spiral density waves in galactic disks are closely
related to each other (Toomre 1981, Bertin et al.~1989a, b). Accordingly
enhanced growth due to the presence of a responsive halo is 
also known for spiral density waves. Mark (1976b) has studied the angular
momentum transfer from mode like spiral density waves to a bulge -- halo
system surrounding the galactic disk. The density waves were modelled as rigidly
rotating exponentially growing spiral patterns whose dynamics was described in
the {\em WKBJ} approximation (Mark 1974, 1976a). The models presented in Mark
(1976b, hereafter referred to as M76) show considerably enhanced growth rates 
of the density waves due to angular momentum loss to the bulge -- halo system.
Recently Fuchs (2004, hereafter referred to as F04) followed a different 
approach to the spiral amplification question in the presence of a live dark
halo. Using the shearing sheet model he showed how a live dark halo
responds to swing amplified spiral density waves and found also a very
enhanced growth of such density waves. These density waves interact with the 
halo via halo particles either on orbits in resonance with the waves or on 
non--resonant orbits.

The shearing sheet model (Goldreich \& Lynden--Bell 1965, Julian \& Toomre 
1966) has been developed as a tool to study the dynamics of a 
patch of a galactic disk. This is assumed to be infinitesimally thin and 
its radial size is assumed to be much smaller than that of the disk. Polar 
coordinates can be therefore rectified to pseudo--Cartesian
coordinates and the velocity field of the differential rotation of the
disk can be approximated by a linear shear flow. These simplifications
allow an analytical treatment of the problem, which helps to clarify 
the underlying physical processes operating in the disk. 

Up to now all studies of the effect of a live dark halo on the dynamics of
non--axisymmetric structures in a galactic disk have assumed an isotropic
velocity distribution of the halo particles. On the other hand, it is well known
from cosmological simulations of dark halo formation that the velocity
distribution of the halo particles must be anisotropic. If the anisotropy is
measured by Binney's parameter $\beta = 1 - (\sigma_\theta^2+\sigma_\phi^2)/2 
\sigma_{\rm r}^2$ (cf.~Binney \& Tremaine 1987) this ranges from
$\beta \approx 0$ near the centers of the halos to $\beta \approx 0.5$ at the
virial radii (C\'olin et al.~2000, Fukushige \& Makino 2001). Hansen \& Moore
(2005) have suggested that the radial variation of $\beta$ is roughly linear to
the logarithmic density gradient of the halo, $\beta \propto -d \ln{\rho}/d
\ln{r} $. Very recently Abadi et al.~(2005) have reported a simulation of a
mildly anisotropic dark halo with an anisotropy parameter of $\beta \approx
0.3$. In the same experiment they have also simulated the formation of the
baryonic galaxy inside the dark halo. Interestingly, the velocity anisotropy  
of the spheroidal component of the model galaxy is significantly larger, 
$\beta \approx 0.4$ to $0.8$, which is consistent with the anisotropy parameter
of $\beta \approx 0.7$ of Galactic halo stars in the solar neighbourhood 
(Arifyanto et al.~2005). Following a different concept, which describes dark
halos as partially relaxed dynamical systems, Trenti \& Bertin (2005) find also
radially anisotropic velocity distributions with an anisotropy parameter in
the region around corotation of $\beta \approx 0.6$. Helmi et al.~(2002)
discuss the implications of such an anisotropy for experimental searches of 
dark matter particles.

Thus the studies of the interaction of dark halos with non--axisymmetric 
structures in galactic disks have to be extended to halo models with 
anisotropic velocity distributions. As a first step we consider in this short 
note the amplification of spiral density waves in disks embedded in
anisotropic dark halos. 

\section{Introducing anisotropic distribution functions for the halo particles}

\subsection{Response of a live dark halo to a density wave in the shearing 
sheet}

We study now the response of a live dark halo with an anisotropic velocity
distribution to a swing amplified density wave in the disk. For this we use the 
method and notation of F04, assuming that the shearing sheet
is immersed in a homogeneous live dark halo. Since we modify only the
unperturbed distribution function of the halo particles in phase space, we can
immediately follow the formalism of F04 to derive the Fourier transformed
Boltzmann equation which describes the evolution of the perturbation of the
distribution function $f_{\rm h {\bf k}}$ (eqn.~20 of F04). We adopt now the
anisotropic background distribution function

\begin{equation}
f_{\rm h0}  = \frac{\rho_{\rm b}}{\sqrt{2 \pi}^3 \sigma_{||}^2
  \sigma_{\rm w}}~ 
{\rm exp} \left( - \frac{u^2+v^2}{2\sigma_{||}^2} - \frac{w^2}
{2\sigma_{\rm w}^2} \right) \,. 
\end{equation}
$u$ and $v$ denote the planar velocity components of the halo particles with 
$v$ pointing in the direction of the differential shear in the sheet. $w$ is the
velocity component perpendicular to the plane of the sheet.
As in F04 we choose orthogonal
Cartesian coordinates $\xi$, $\eta$, and $\zeta$ with $\xi$ parallel to the
wave vector ${\bf k}$ of $f_{\rm h {\bf k}}$. According to the 
symmetry of the distribution function (1), one of the coordinates perpendicular
to ${\bf k}$ can be aligned without loss of generality with the $v$ velocity 
component. Thus the distribution function (1) can be written as

\begin{eqnarray}
&& f_{\rm h0} = 
\frac{\rho_{\rm b}}{\sqrt{2 \pi}^3 \sigma_{||}^2 \sigma_{\rm w}}\,
{\rm exp} \left( - \frac{v^2}{2\sigma_{||}^2} -  
\frac{(v_{||}\cos{\alpha}-v_\perp \sin{\alpha})^2}{2\sigma_{||}^2}
\right) \nonumber \\ 
&& \times {\rm exp} \left( -
\frac{(v_{||}\sin{\alpha}+v_\perp \cos{\alpha})^2}{2\sigma_{\rm w}^2}
\right)  \,,
\end{eqnarray}

\noindent
where $\cos{\alpha}= k_{||}/\sqrt{k_{\rm z}^2 + k_{||}^2}$, and $k_{||}=
|{\bf k}_{||}|=\sqrt{k_{\rm x}^2+k_{\rm y}^2}$ is the component of the wave 
vector ${\bf k}$ parallel to the mid-plane of the sheet. $v_\perp$ denotes 
the second velocity component perpendicular to ${\bf k}$. The Boltzmann equation
can be then immediately integrated with respect to $v$ and, after some
algebra, the integration with respect to $v_\perp$ leads to a Boltzmann
equation which has the same form as eqn.~(20) of F04, but where the
one--dimensional velocity dispersion $\sigma_{\rm h}$ of the isotropic 
distribution function is replaced by the velocity dispersion of the velocity 
component in the $\xi$--direction parallel to ${\bf k}$

\begin{equation}
\sigma_{\rm eff} = \sqrt{\sigma_{||}^2\cos{\alpha}^2 +
\sigma_{\rm w}^2\sin{\alpha}^2} \,. 
\end{equation}

\noindent
We can proceed then exactly as in F04 and obtain the same final formal result
$\Phi_{\rm h{\bf k}_{||}}(z=0) =\Upsilon (k_{\rm x}, k_{\rm y}, \omega)
 \Phi_{\rm d{\bf k}_{||}}$,
where the function $\Upsilon$ is modified by replacing $\sigma_{\rm h}$ by
$\sigma_{\rm eff}$ in eqns.~(15) and (17) of F04. 

As alternatives to the distribution function of the halo particles (1) we have 
considered also the anisotropic distribution functions
 
\begin{equation}
f_{\rm h0}  = \frac{\rho_{\rm b}}{\sqrt{2 \pi}^3 \sigma_{\rm u}
  \sigma_{\rm w}^2}~ 
{\rm exp} \left( - \frac{u^2}{2\sigma_{\rm u}^2} - \frac{v^2+w^2}
{2\sigma_{\rm w}^2} \right)  
\end{equation}

\noindent
and

\begin{equation}
f_{\rm h0}  = \frac{\rho_{\rm b}}{\sqrt{2 \pi}^3 \sigma_{\rm v}
  \sigma_{\rm w}^2}~ 
{\rm exp} \left( - \frac{v^2}{2\sigma_{\rm v}^2} - \frac{u^2+w^2}
{2\sigma_{\rm w}^2} \right)\,.  
\end{equation}

\noindent
If these are integrated with respect to the two velocity components
perpendicular to {\bf k}, we find the effective velocity dispersions

\begin{equation}
\sigma_{\rm eff} = \sqrt{\sigma_{\rm u}^2\cos{\beta}^2 +
\sigma_{\rm w}^2\sin{\beta}^2} 
\end{equation}

\noindent
and 

\begin{equation}
\sigma_{\rm eff} = \sqrt{\sigma_{\rm v}^2\cos{\gamma}^2 +
\sigma_{\rm w}^2\sin{\gamma}^2}\,, 
\end{equation} 

\noindent
respectively. The angles $\beta$ and $\gamma$ are given by $\cos{\beta}$ = 
$k_{\rm x}/\sqrt{k_{\rm z}^2+k_{||}^2}$ and $\cos{\gamma}$ = 
$k_{\rm y}/\sqrt{k_{\rm z}^2+k_{||}^2}$. An effective velocity
dispersion $\sigma_{\rm eff}$ can also be calculated analytically for a 
triaxial distribution function. However, the result is so cumbersome that we do
not consider this case here.

\subsection{Dynamics of the shearing sheet embedded in an anisotropic
 live dark halo} 

The halo response to the perturbation in the disk has to be inserted into
the Boltzmann equation describing the evolution of the distribution function of
the disk particles in phase space. We follow again the formalism of F04 and 
find the fundamental Volterra type integral equation for the
Fourier transforms of the perturbations of the gravitational potential of the
shearing sheet in the same form as eqn.~(33) of F04, but with a modified Fourier
transform $\mathcal{F} ( \Upsilon e^{i\omega\frac{k_{\rm x}-k'_{\rm x}}
{2 A k'_{\rm y}}} )_{\rm t-t'}$. The latter can be calculated analytically as in
F04 using first formulae 6.317 and 3.952 of Gradshteyn \& Ryzhik (2000) and 
then formulae 3.223 and 3.466 for the integrals with respect to $k_{\rm z}$, 
leading to

\begin{eqnarray}
&& \mathcal{F}\left(\Upsilon(k_{\rm x},k'_{\rm y},\omega)
e^{i\omega\frac{k_{\rm x}-k'_{\rm x}}{2 A k'_{\rm y}}}\right)_{\rm t-t'} = 
\nonumber \\ && \frac{8 \pi^2 G \rho_{\rm b}}
{\sigma_{||}(\sigma_{||}+\sigma_{\rm w})}\frac{1}{k_{\rm
x}^2+{k'_{\rm y}}^2} 
~\delta \left( t-t'+ \frac{k_{\rm x}-k'_{\rm x}}{2 A k'_{\rm y}} \right) \\ &&
+ 4 \pi^2G \rho_{\rm b}\exp {\left [ik'_{\rm y}r_{\rm 0}\Omega_{\rm 0}
\left( t-t'+ \frac{k_{\rm x}-k'_{\rm x}}{2 A k'_{\rm y}} \right)
 \right]} \nonumber \\ && \times
\left \{ \left( t-t'+ \frac{k_{\rm x}-k'_{\rm x}}{2 A k'_{\rm y}} \right) + 
\Big|
t-t'+ \frac{k_{\rm x}-k'_{\rm x}}{2 A k'_{\rm y}} \Big| \right \} \nonumber \\
 && \times \exp{\big[ -\frac{1}{2} (\sigma_{||}^2-\sigma_{\rm w}^2)
(k_{\rm x}^2 + {k'}_{\rm x}^2) \left( t-t'+ \frac{k_{\rm x}-k'_{\rm x}}
{2 A k'_{\rm y}} \right)^2\big]}    \nonumber \\ &&
\times {\rm erfc} \left( \frac{\sigma_{\rm w}}{\sqrt{2}} \sqrt{k_{\rm x}^2 
+{k'_{\rm y}}^2}~\Big|t-t'+ \frac{k_{\rm x}-k'_{\rm x}}{2 A k'_{\rm y}} 
\Big| \right)\nonumber \,,
\end{eqnarray}

\noindent
The modified equation (35) of F04 can be integrated numerically still with very
modest numerical effort. The characteristic response of the shearing sheet
embedded in a live dark halo to an initial impulse is not changed qualitatively
by the anisotropy of the velocity distribution of the halo particles. The 
shearing sheet develops swing amplified density waves with their amplitudes 
enhanced by the presence of a live dark halo as compared to a static halo
(cf.~Fig.~2 of F04). In the case of the distribution function (4) we obtain the
Fourier transform

\begin{eqnarray}
&& \mathcal{F}\left(\Upsilon(k_{\rm x},k'_{\rm y},\omega)
e^{i\omega\frac{k_{\rm x}-k'_{\rm x}}{2 A k'_{\rm y}}}\right)_{\rm t-t'} = 
\nonumber \\ && \frac{8 \pi^2 G \rho_{\rm b}}
{\sqrt{\sigma_{\rm u}^2 k_{\rm x}^2+\sigma_{\rm w}^2 {k'}_{\rm y}^2}
(\sigma_{\rm w} \sqrt{k_{\rm x}^2+{k'}_{\rm y}^2} +
\sqrt{\sigma_{\rm u}^2 k_{\rm x}^2+\sigma_{\rm w}^2 {k'}_{\rm y}^2})}
\nonumber \\ && \times
\delta \left( t-t'+ \frac{k_{\rm x}-k'_{\rm x}}{2 A k'_{\rm y}} \right) \\ &&
+ 4 \pi^2 G \rho_{\rm b}\exp {\left [ik'_{\rm y}r_{\rm 0}\Omega_{\rm 0}
\left( t-t'+ \frac{k_{\rm x}-k'_{\rm x}}{2 A k'_{\rm y}} \right)
 \right]} \nonumber \\ && \times
\left \{ \left( t-t'+ \frac{k_{\rm x}-k'_{\rm x}}{2 A k'_{\rm y}} \right) + 
\Big|
t-t'+ \frac{k_{\rm x}-k'_{\rm x}}{2 A k'_{\rm y}} \Big| \right \} \nonumber \\
 && \times \exp{\big[ -\frac{1}{2} (\sigma_{\rm u}^2-\sigma_{\rm w}^2)
k_{\rm x}^2 \left( t-t'+ \frac{k_{\rm x}-k'_{\rm x}}
{2 A k'_{\rm y}} \right)^2\big]}    \nonumber \\ &&
\times {\rm erfc} \left( \frac{\sigma_{\rm w}}{\sqrt{2}} \sqrt{k_{\rm x}^2 
+{k'_{\rm y}}^2}~\Big|t-t'+ \frac{k_{\rm x}-k'_{\rm x}}{2 A k'_{\rm y}} 
\Big| \right)\nonumber \,,
\end{eqnarray}

\noindent
and similarly for the distribution function (5)

\begin{eqnarray}
&& \mathcal{F}\left(\Upsilon(k_{\rm x},k'_{\rm y},\omega)
e^{i\omega\frac{k_{\rm x}-k'_{\rm x}}{2 A k'_{\rm y}}}\right)_{\rm t-t'} = 
\nonumber \\ && \frac{8 \pi^2 G \rho_{\rm b}}
{\sqrt{\sigma_{\rm w}^2 k_{\rm x}^2+\sigma_{\rm v}^2 {k'}_{\rm y}^2}
(\sigma_{\rm w} \sqrt{k_{\rm x}^2+{k'}_{\rm y}^2} +
\sqrt{\sigma_{\rm w}^2 k_{\rm x}^2+\sigma_{\rm v}^2 {k'}_{\rm y}^2})}
\nonumber \\ && \times
\delta \left( t-t'+ \frac{k_{\rm x}-k'_{\rm x}}{2 A k'_{\rm y}} \right) \\ &&
+ 4 \pi^2 G \rho_{\rm b}\exp {\left [ik'_{\rm y}r_{\rm 0}\Omega_{\rm 0}
\left( t-t'+ \frac{k_{\rm x}-k'_{\rm x}}{2 A k'_{\rm y}} \right)
 \right]} \nonumber \\ && \times
\left \{ \left( t-t'+ \frac{k_{\rm x}-k'_{\rm x}}{2 A k'_{\rm y}} \right) + 
\Big|
t-t'+ \frac{k_{\rm x}-k'_{\rm x}}{2 A k'_{\rm y}} \Big| \right \} \nonumber \\
 && \times \exp{\big[ -\frac{1}{2} (\sigma_{\rm v}^2-\sigma_{\rm w}^2)
{k'}_{\rm y}^2 \left( t-t'+ \frac{k_{\rm x}-k'_{\rm x}}
{2 A k'_{\rm y}} \right)^2\big]}    \nonumber \\ &&
\times {\rm erfc} \left( \frac{\sigma_{\rm w}}{\sqrt{2}} \sqrt{k_{\rm x}^2 
+{k'_{\rm y}}^2}~\Big|t-t'+ \frac{k_{\rm x}-k'_{\rm x}}{2 A k'_{\rm y}} 
\Big| \right)\nonumber \,.
\end{eqnarray}

\noindent
The characteristic response of the shearing sheet embedded in dark halos with
these anisotropic distribution functions of the halo particles to an initial 
impulse is still the same as in the previous case, only the maximum growth
factors of the density waves change. All three Fourier transforms (8,
9, and 10) are reduced to the form given in F04, if the distribution functions
become isotropic.

\subsection{Numerical applications}

\subsubsection{Comparison with previous work}

No results with anisotropic halos, either analytic or numerical, have
been published so far. Indeed, as mentioned already in the introduction, 
the lack of such information has been one of the main motivations for our 
work. A couple of previous works, however, have studied 
the influence of a live dark halo with an isotropic distribution function on
density waves and we will present here some possible, albeit rough, comparisons. 

The interaction of a {\em WKBJ} type density wave with a separate
non--rotating component was studied for the first time by Marochnik \& Suchkov
(1969). The astronomical community was at that time not aware of the 
existence of dark halos, but the considerations of Marochnik \& Suchkov (1969)
carry over to the disk -- dark halo interaction. Unfortunately, only very few
concrete details were given. Much more detailed is the investigation by Mark 
(M76). In Fig.~3 of M76 it is shown how much stronger the amplification of the 
density waves is
in the presence of a live halo than in an isolated disk. This depends on
the assumed corotation radii of the spiral patterns. Since swing amplified
density waves are basically corotating with the center of the shearing sheet, we
adopt for each corotation radius of the spiral modes the parameters given in 
Table 1 of M76 for that galactocentric radius and calculate for these parameters
the corresponding maximum growth factors of swing amplified density waves in the
shearing sheet. The regions in the galaxy models, where the rotation curves 
are essentially flat, are most suitable
for a comparison. Using the formalism discussed in the previous
sections, we find for the parameters of model A$_1$ at R = 6 kpc 
an enhancement of the maximum growth factor of the amplitudes of swing 
amplified density waves due to the halo by a factor of 2, while Fig.~3 of M76
indicates that the growth rates of the density waves are enhanced by a
factor of 2.3. Similarly, we find for the
parameters of model B$_1$ in the range R = 14 to 18 kpc maximum growth factors
enhanced by factors of 3.2 to 1.4. Fig.~3 of M76 indicates for that
range an enhancement of the growth rates by a factor of 1.7. Thus the 
numerical values are in rough agreement.

\subsubsection{Numerical applications to anisotropic halos}

\begin{figure}
\rotatebox{-90}{\resizebox{0.7\hsize}{!}{\includegraphics 
{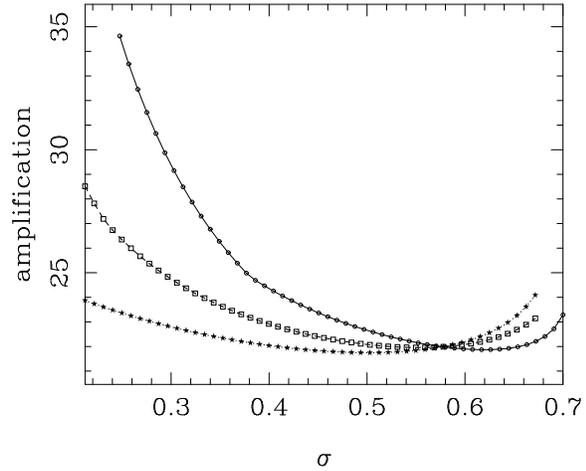}}}
\caption{ Maximum growth factor of the amplitudes of density waves in the
shearing sheet as function of the anisotropy of the velocity distribution
of the halo particles. The solid line with open circles illustrates the case if
the components of the velocity dispersion are varied as $\sigma_{\rm u} = 
\sigma_{\rm v}$. These are drawn as abscissa. The third component is chosen as 
$\sigma_{\rm w} = \sqrt{1-2 \sigma_{\rm u}^2 }$, so that the total velocity 
dispersion is constant. The cases $\sigma_{\rm v} = \sigma_{\rm w}$,
$\sigma_{\rm u} = \sqrt{1-2 \sigma_{\rm v}^2 }$ are shown as the dashed line
with open squares and $\sigma_{\rm u} = \sigma_{\rm w}$,
$\sigma_{\rm v} = \sqrt{1-2 \sigma_{\rm u}^2 }$ as the dotted line with filled
asterisks, respectively. The isotropic case $\sigma=$ 0.58 reproduces 
the result of F04.}
\label{fig:1}       
\end{figure}

We have determined maximum growth factors of the amplitudes of the density 
waves for various values of the parameters characterizing the disk -- halo
configuration.
Representative results for the three cases described above are
presented in Fig.~1. In all three cases we have taken
$\sigma_{\rm tot}^2~=~\sigma_{\rm u}^2~+~\sigma_{\rm v}^2~+
~\sigma_{\rm w}^2~=~1$, $A / \Omega_0$ = 0.5 and $k'_{\rm y}$
 = 0.5 $k_{\rm crit}$. In F04 it was shown
how equations (8) to (10) can be cast into dimensionless form. The velocity
dispersion of the halo particles is then measured in terms of the velocity
dispersion of the disk stars. We have assumed for the examples in Fig.~1
a ratio of $\sigma_{\rm tot}:\sigma_{\rm disk} = 5 : 1$. The velocity 
dispersion of the disk stars is determined by the Toomre stability parameter 
for which we have adopted a value of $Q$ = 1.41. For the density of the dark
halo and the circular velocity of the center of the shearing sheet around the
galactic center we have adopted values typical for the solar neighbourhood
in the Milky Way, namely $\rho_{\rm b}G/\kappa^2$ = 0.01 and $r_{0}\Omega_0$ :
$\sigma_{\rm disk}$ = 220 : 44.
We find in all three cases the same characteristic behaviour, that the
amplification of density waves in the shearing sheet is smallest, if the 
distribution function of the halo particles is almost, but not exactly,
isotropic. Even in the isotropic case, though, the amplification is
still considerably larger than in a shearing sheet  
embedded in a static halo. The dependence of the maximum growth factor on the
velocity dispersion components is rather subtle. If we concentrate on the 
case $ \sigma_{\rm u}=\sigma_{\rm v}$, it becomes clear 
from equation (8) that
$\mathcal{F}$ rises, if $\sigma_{||}$ is decreasing with respect to 
$\sigma_{\rm w}$, which leads to a stronger input to the swing amplification 
mechanism. This can be intuitively understood as the halo becomes more 
susceptible to the sheared trailing spiral arms in the disk, if the planar
velocity dispersions become small. The dependence of
$\mathcal{F}$ on the $\sigma_{\rm w}$ velocity dispersion component is twofold.
Technically, this is related to the integrations of the halo response to the 
disk
perturbations with respect to the $k_{\rm z}$ wave number (cf.~eqns.~26 and 28
of F04). The result is that $\mathcal{F}$ rises also at lower $\sigma_{\rm w}$
velocity dispersions. This means that the dark halo becomes then less stiff in
the vertical direction so that it can support the density waves in the disk more
effectively, although the effect is less pronounced. The relative minimum 
of amplification of density waves is
at slightly larger $\sigma_{\rm u}=\sigma_{\rm v}$ velocity dispersions than the
isotropic case $\sigma_{\rm u}=\sigma_{\rm v}=\sigma_{\rm w}=0.58\,
\sigma_{\rm tot}$. The other diagrams in Fig.~1 can be explained by similar 
arguments. However, since in these cases only one planar velocity
dispersion is varied, the maximum growth factor is less than in the previous 
case. We have also considered other numerical values for the
parameters $Q$ and $A$ of the disk as well as for the halo
parameters $\rho_{\rm b}$ and $\sigma_{\rm tot}$. Even though the overall
amplification of the density waves can change drastically, the effect of the
anisotropy as compared to the isotopic case remains always roughly the same.

Extrapolating these results, we predict that spirals and bars growing in
disk galaxies with non--isotropic halos should be on average stronger than
those in isotropic halos. Unfortunately it is not possible to test
this for real galaxies. It is, however, possible to do so for $N$--body
simulations and such a study is presently underway (Athanassoula, in prep.). 

\acknowledgements{We thank the anonymous referee for his comments which helped 
to improve the paper.}

{}


\begin{thebibliography}{}

   \bibitem[2005]{abadi} Abadi, M.G., Navarro, J., Steinmetz, M., 2005,
   astr-ph/0506659
   
   \bibitem[2005]{arif} Arifyanto, M. I., Fuchs, B., Jahrei{\ss}, H., 
   Wielen, R., 2005, A\&A, 433, 911
   
   \bibitem[1996]{Ath96} Athanassoula, E., 1996, in: 
   R. Buta, D.A. Crocker, B.G. Elmegreen (eds.) Barred Galaxies, 
   Astron. Soc. Pac. Conf. Series, 91, p. 309

   \bibitem[2002]{ath02} Athanassoula, E., 2002, ApJ, 569, L83 
   
   \bibitem[2003]{ath03} Athanassoula, E., 2003, MNRAS, 341, 1179
   
   \bibitem[1987]{bintre} Binney, J., Tremaine, S., 1987, Galactic Dynamics,
   Princeton University Press, Princeton
   
   \bibitem[1989]{ber} Bertin, G., Lin, C.C., Lowe, S.A., Thurstans, R.P.,
   1989a, ApJ, 338, 78
   
   \bibitem[1989]{berli} Bertin, G., Lin, C.C., Lowe, S.A., Thurstans, R.P.,
   1989b, ApJ, 338, 104
   
   \bibitem[1989]{coli} C\'olin, P., Klypin, A.A., Kravtsov, A.V., 2000, ApJ, 
   539, 561
   
   \bibitem[2000]{deba} Debattista, V.P., Sellwood, J.A., 2000, ApJ, 543, 704 
   
   \bibitem[2004]{fuchs04} Fuchs, B., 2004, A\&A, 419, 941

   \bibitem[2001]{fuku} Fukushige, T., Makino, J., 2001, 557, 533
   
   \bibitem[1965]{gold} Goldreich, P., Lynden--Bell, D., 1965, MNRAS, 
   130, 125
   
   \bibitem[1965]{grad} Gradshteyn, I.S., Ryzhik, I.M., 2000, Table of 
   Integrals, Series, and Products (6th Ed.), Academic Press, New York
   
   \bibitem [2005]{Hanmo} Hansen, S. , Moore, B., 2005, astro-ph/04011473
   
   \bibitem [2005]{Helmi} Helmi, A., White, S.D.M., Springel, V., 2002, PRD, 66,
   063502
   
   \bibitem [1992]{HW92} Hernquist, L., Weinberg, M., 1992, ApJ, 400,
   80

   \bibitem[1966]{julia} Julian, W.H., Toomre, A., 1966, ApJ, 146, 810

   \bibitem[1971]{kal} Kalnajs, A. J., 1971, ApJ, 166, 275
  
   \bibitem[1991]{LC2} Little, B., Carlberg, R.G., 1991, MNRAS, 251, 227

   \bibitem[1972]{lbk} Lynden-Bell, D., Kalnajs, A.J., 1972, MNRAS,
   157, 1
    
   \bibitem[1974]{mar} Mark, J.W.K., 1974, ApJ, 193, 539
   
   \bibitem[1976]{mark} Mark, J.W.K., 1976a, ApJ, 205, 363
   
   \bibitem[1976]{markj} Mark, J.W.K., 1976b, ApJ, 206, 418
   
   \bibitem[1969]{maro} Marochnik, L.S., Suchkov, A.A., 1969, AZ, 46, 319

   \bibitem[2003]{OND} O'Neill, J.K., Dubinski, J., 2003, MNRAS, 346, 251 
   
   \bibitem[1977]{t77} Toomre, A., 1977, ARAA, 15, 437
   
   \bibitem[1981]{too} Toomre, A., 1981, in: S.M. Fall, D. Lynden--Bell (eds.)
    The Structure and Evolution of Normal Galaxies, Cambridge Univ. Press, 
    Cambridge, p. 111 

   \bibitem[1984]{tremwein} Tremaine, S., Weinberg, M. D., 1984, MNRAS,
   209, 729
   
   \bibitem[1984]{trenti} Trenti, M., Bertin, G., 2005, A\&A, 429, 161
   
   \bibitem[2003]{valkly} Valenzuela, O., Klypin, A., 2003, MNRAS, 345, 406

   \bibitem[1985]{wei} Weinberg, M.D., 1985, MNRAS, 213, 451

\end{thebibliography}
\end{document}